\documentclass[a4paper,11pt]{article}
\pdfoutput=1 

\usepackage{jinstpub} 

\usepackage{tikz}
\tikzset{
  font={\fontsize{8pt}{10}\selectfont}}
  
\title{\boldmath APRIL : a novel Algorithm for Particle Reconstruction at ILC}

\author{B. Li,}
\author[1]{R. \'Et\'e,\note{Now at DESY.}}
\author[2]{G. Grenier,\note{Corresponding author.}}
\author{and I. Laktineh}

\affiliation{IP2I Lyon, Universit\'e Lyon 1, CNRS/IN2P3, Universit\'e de Lyon,\\ 4 rue Enrico Fermi, 69622 Villeurbanne CEDEX, France}

\emailAdd{g.grenier@ip2i.in2p3.fr}

\abstract{
The current developments for future electron-positron colliders are driven by the Particle Flow concept. 
In these developments, high granularity calorimeters play a central role. This presentation will focus
on a new Particle Flow Algorithm (PFA) developed for high granularity calorimeters, and especially for
the Semi-Digital Hadronic CALorimeter (SDHCAL) option of the International Large Detector (ILD) concept. 
The first PFA for ILD was PandoraPFA. This new PFA (APRIL) is based on the PandoraPFA Software Development Kit,
but implements a different clustering inspired from the ARBOR PFA approach. This proceeding will describe 
briefly the APRIL algorithm and discuss its performance compared to that of PandoraPFA.  

}


\arxivnumber{2002.09678} 

\proceeding{3$^{\text{rd}}$ conference on Calorimetry for the High Energy Frontier (CHEF)\\
  November 25-29, 2019\\
  Kyushu University, Fukuoka, Japan}

\begin{document}
\maketitle
\flushbottom

\section{Particle flow calorimetry}
\label{sec:introPFA}

The Particle Flow reconstruction, pioneered in \cite{Brient:2002gh,Morgunov:2002pe}, consists in measuring jets by following each jet particle in the full detector and using the most suitable sub-detector to measure the particle energy.
The jet energy is then measured as the sum of the charged particles energy measured in the tracker, the photons energy measured in the electromagnetic calorimeter (ECAL) and the neutral hadrons energy measured in the hadronic calorimeter (HCAL). 

The particle flow approach is a key ingredient for ILC detectors to reach the design target of better than $3-4$\% jet energy resolution for jets in the 50~GeV to 500~GeV range \cite{Behnke:2013lya}. 
The implementation of Particle Flow Algorithm (PFA) needs highly granular calorimeters. For example, for the International Large Detector (ILD) concept \cite{Behnke:2013lya}, the base option for ECAL is a Silicon-Tungsten sampling calorimeter (SiWECAL) with 29 layers, 2 to 4 mm thick with $5\times 5$~mm$^2$ detection cells. 
The ILD hadronic calorimeter would be an iron sampling calorimeter with 48 layers, 25 to 26 mm thick with either a first option with an analogue readout of $30 \times 30$~mm$^2$ scintillator cells (AHCAL) or a second option with a semi-digital readout of $10 \times 10$~mm$^2$ Glass Resistive Plate Chamber (GRPC) cells (SDHCAL).
All these technologies have been developed within the CALICE collaboration and the corresponding prototypes are described in Ref. \cite{Anduze:2008hq} for the SiWECAL, Ref. \cite{collaboration:2010hb} for the AHCAL and Ref. \cite{Baulieu:2015pfa} for the SDHCAL.

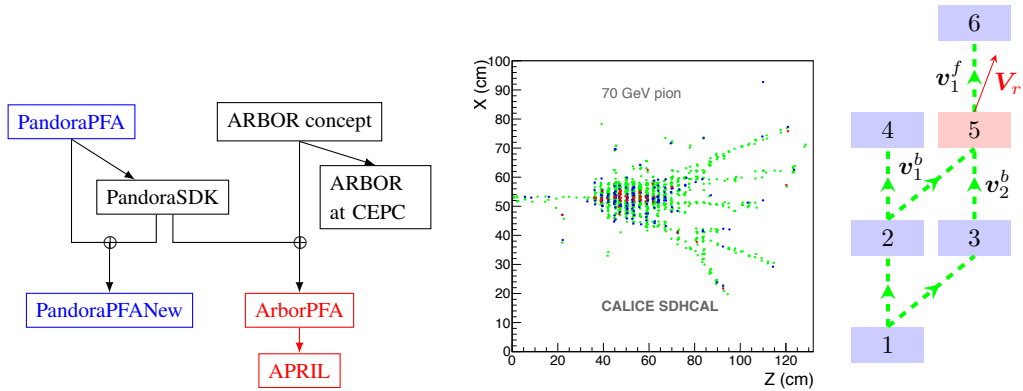
\begin{figure}[htbp]
\centering 
\begin{minipage}[t]{0.4\textwidth}
\begin{tikzpicture}
\node[draw,blue] (pandora) at (0,0) {PandoraPFA};
\node[draw] (arbor) at (3,0) {ARBOR concept};
\node[draw] (pandorasdk) at (1.2,-1) {PandoraSDK};
\node[draw,text width=1.2cm] (cepc) at (4,-1) {ARBOR at CEPC};
\node (pluspandora) at (0.5,-1.6) {$\oplus$};
\node (plusarbor) at (3,-1.6) {$\oplus$};
\node[draw,blue] (pandoranew) at (0.5,-2.5) {PandoraPFANew}; 
\node[draw,red] (arborpfa) at (3,-2.5) {ArborPFA}; 
\node[draw,red] (april) at (3,-3.3) {APRIL};
\draw[->,>=latex] (pandora.south) -- (pandorasdk);
\draw[->,>=latex] (arbor.south) -- (cepc.north);
\draw (pandora.south) |- (pluspandora.center);
\draw (pandorasdk.250) |- (pluspandora.center);
\draw (pandorasdk.300) |- (plusarbor.center);
\draw (arbor) -- (plusarbor.center);
\draw[->,>=latex] (pluspandora.center) -- (pandoranew);
\draw[->,>=latex] (plusarbor.center) -- (arborpfa);
\draw[->,>=latex,red] (arborpfa) -- (april);
\end{tikzpicture}
\end{minipage}
\includegraphics[width=.3\textwidth]{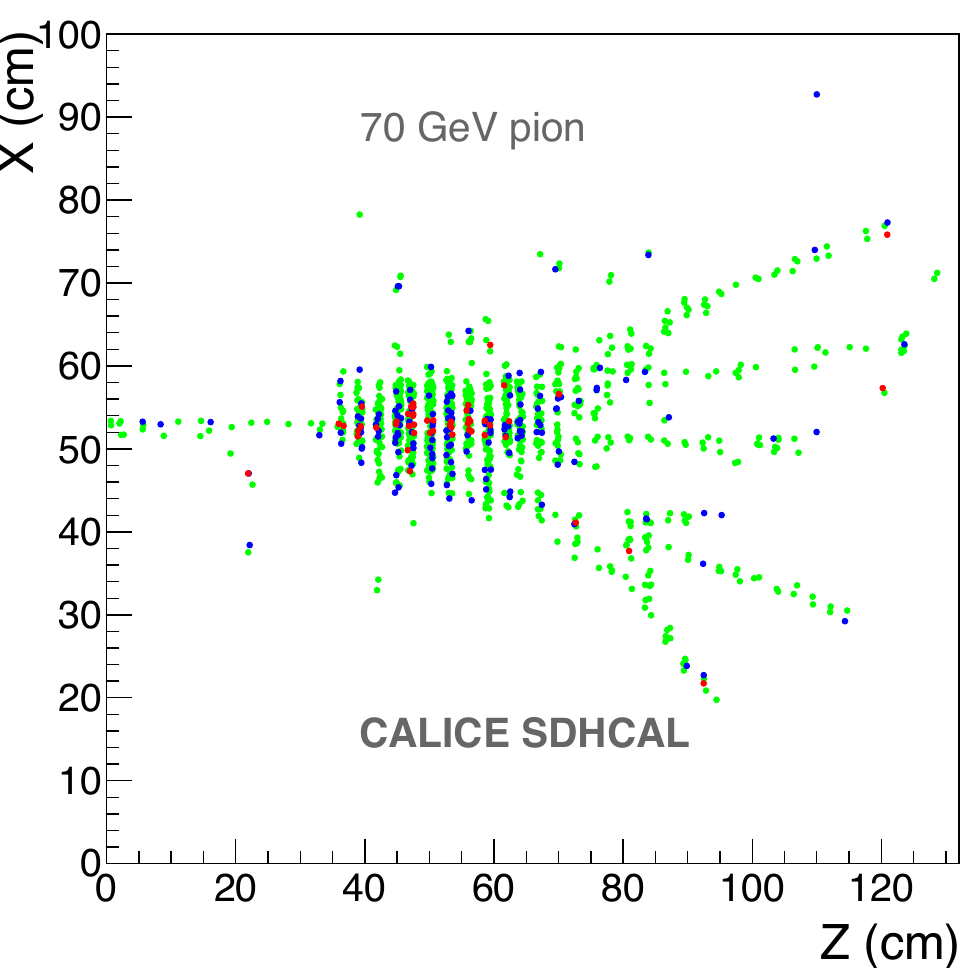}
\includegraphics[width=.2\textwidth]{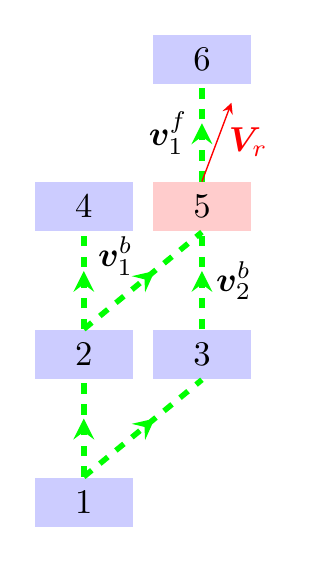}
\caption{\label{fig:i} Left : PFA reconstruction techniques used for ILD studies. 
Middle : An event display of a 70 GeV pion shower recorded in the CALICE SDHCAL prototype. Colour points indicate the higher crossed threshold: first (green), second (blue) and third (red).
Right: An example of a 2D projection of the initial ARBOR clustering of a group of 6 hits. For hit number 5, the reference direction used in the connector/edge cleaning step is shown. }
\end{figure}

In parallel to the design of the calorimeter, the software development of PFA has been pursued. Figure \ref{fig:i} left shows the evolution of PFA developed for the ILD. The first PFA developed had been PandoraPFA \cite{Thomson:2009rp} which have been rewritten and separated in two pieces, a software toolkit PandoraSDK \cite{Marshall:2015rfa} and the Pandora PFA clustering method (PandoraPFANew). The PandoraPFA has been developed for the ILD option 1, so assuming purely analogue calorimeters.
In parallel, a separate approach for the ILD option 2, ARBOR  \cite{Ruan:2014paa} has been developed. This concept has been implemented as the ARBOR PFA for CEPC physics studies \cite{Ruan:2018yrh} and for ILD using the PandoraSDK \cite{Ete:2017rym,Ete:2015npj}. This latter implementation is currently being upgraded to a new version named APRIL (A Particle Reconstruction for ILC - Lyon). 
This proceeding will briefly describe the APRIL algorithm and discuss its performance compared to that of PandoraPFA.

\section{The SDHCAL energy reconstruction}
\label{sec:SDHCAL}

When a particle shower develops in the SDHCAL, each charged particle entering the GRPC gas gap produces an electrons avalanche. 
These avalanches induce an electric signal on $10 \times 10$~mm$^2$ copper pads read out by an embedded electronics producing a 2-bit output per pad. The output value is set depending on the biggest threshold crossed by the signal. For the SDHCAL prototype, the 3 threshold values are set to 0.11, 5 and 15~pC corresponding approximately to one avalanche, a few avalanches and many avalanches occurring in front of the pad. An event display of a recorded shower is shown on Figure \ref{fig:i} middle.

The SDHCAL energy reconstruction is done by counting the number of hits produced by the shower:
\begin{equation}
\label{eq:Ereco}
 E_{reco}= \alpha_1 N_1 + \alpha_2 N_2 + \alpha_3 N_3 ,
\end{equation}
where $N_i$ is the number of hits for which $i$ is the highest crossed threshold. Few options have been used for the $\alpha_i$ coefficient. 
In test beam data, $\alpha_i$ are quadratic function of the total number of hits in the shower. 
For ILD reconstruction, $\alpha_i$ are constant so that Eq. \ref{eq:Ereco} is linear and thus compatible with PandoraPFA reconstruction. In addition, the $\alpha_i$ parameters' value can be modulated by the number of surrounding hits to treat differently the core and exterior of a shower. 

Using constant values for $\alpha_i$, the PandoraPFA algorithm works with the SDHCAL. 
ArborPFA has originally been developed with the same enrgy reconstruction both for simplicity and to be able to compare its performance against that of PandoraPFA.
In both PFA, the strategy is to construct many small clusters and then to merge them to match each track momentum with associated merged clusters energy. 
The first implementation of ArborPFA, though operational, was not reaching the PandoraPFA performance. A careful study has spotted that the cause was in the cluster merging procedure. 
The APRIL PFA is then the ArborPFA clustering with a revamped cluster merging. 

\section{The APRIL algorithm}
\label{sec:APRIL}

\begin{figure}[htbp]
\centering 
\includegraphics[width=1.\textwidth]{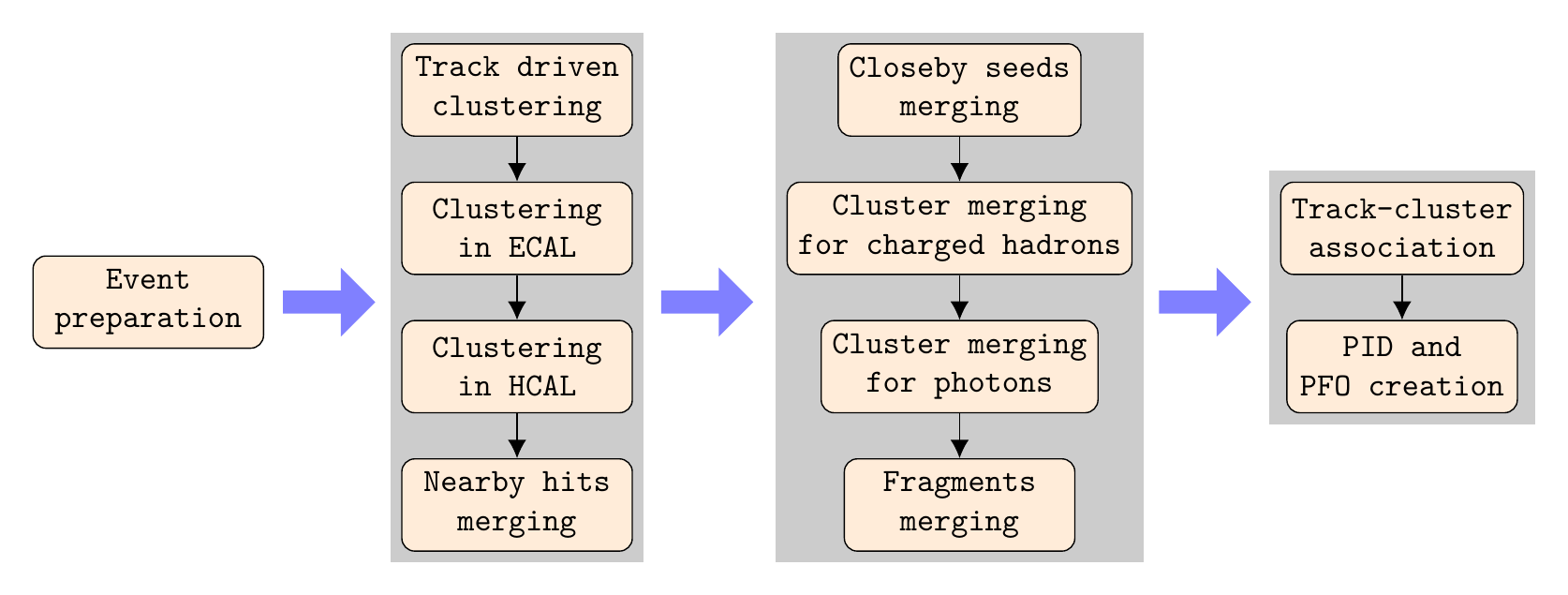}
\caption{\label{fig:iii} The APRIL algorithm layout, The four  sequences are event preparation, clustering, cluster merging and reconstructed particle (PFO) creation.}
\end{figure}

The APRIL algorithm is implemented in the PandoraSDK environment. 
Its main steps are described on Figure \ref{fig:iii}. These steps are grouped in 4 sequences : event preparation, clustering, cluster merging and reconstructed particle (PFO) creation. The two middle sequences, clustering and cluster merging are the prime components of APRIL and will be described in subsections \ref{subsec:ARBOR} and \ref{subsec:ClusterMerging}. 

The initialisation sequence, event preparation, consists in converting hits and tracks into PandoraSDK C++ data types and grouping the hits by subdetector (ECAL, HCAL, ...).  
The finalisation sequence is done in two steps. 
First, tracks and clusters are grouped into PandoraSDK C++ objects, named Particle Flow Object (PFO). Then each PFO is assigned one of the 3 following labels, charged particle if it contains a track, photon or neutral hadron. A PFO object represents a reconstructed particle.

\subsection{ARBOR clustering}
\label{subsec:ARBOR}

The APRIL clustering sequence is the following:
\begin{itemize}
\item Track driven clustering: extrapolate tracks in the calorimeters. The hits close to a track extrapolation are labelled as track connected.
\item Perform the clustering using the ARBOR method on all the hits separately in ECAL and HCAL. Clusters containing track connected hits are identified as charged clusters.
\item Nearby hits merging: remaining non-clustered hits are clustered with the mlpack DBSCAN algorithm \cite{mlpack2018}, an efficient nearest neighbour clustering algorithm.  
\end{itemize} 

The ARBOR method for clustering is based on graph theory. A calorimeter shower is seen as an oriented tree.
The forward direction for the tree is defined as the outgoing direction from the initial collision point at the detector center. This direction is measured in units of interaction length.
Since the ILD calorimeters are sampling calorimeters, hits are grouped 
in virtual nested cylinders, called pseudo-layers. 
These cylinders are deformed so that hits belonging to the same pseudo-layer have roughly the same number of interaction lengths from the detector center. A hit will be considered to be more forward if its pseudo-layer number is bigger.

Each hit is a node for the graph. Initial graph edges are made by connecting each hit with its neighbour hits, determined using the mlpack NeighborSearch function \cite{mlpack2018}.  
Then, for each hit, only one backward edge is kept. The kept edge is the one minimizing a figure of merit $\kappa$ where:
\begin{subequations}
\begin{align}
\label{eq:connectorcleaning:kappa}
\kappa &= \theta^{p_{\theta}} \times d^{p_{d}} ,
\\
\label{eq:connectorcleaning:vr}
\mathbf{V}_{r} &= w_{b} \times \sum_{i} \mathbf{v}_{i}^{b} + w_{f} \times \sum_{j} \mathbf{v}_{j}^{f} ,
\end{align}
\end{subequations}
where $\theta$ is the angle between the edge and the reference direction $\mathbf{V}_{r}$, $d$ is the distance, in radiation length in ECAL or interaction length in HCAL, between the hits connected by the edge and $p_{\theta}$ and $p_{d}$ are parameters. 
The reference direction, illustrated on Figure \ref{fig:i} right, is computed with Eq. \ref{eq:connectorcleaning:vr} where $\mathbf{v}_{i}^{b}$ and $\mathbf{v}_{j}^{f}$ are the backward edge $i$ direction  and forward edge $j$ direction. By adjusting the depth of the edge summation and the forward and backward weights $w_{f}$ and $w_{b}$, the tree slenderness can be adjusted.

\subsection{Cluster merging}
\label{subsec:ClusterMerging}

The APRIL cluster merging is done by applying the ARBOR method to the clusters. As the first step, all charged clusters are connected with their neighbouring neutral clusters. 
For each neutral cluster, a maximum of one connection with a charged  cluster is kept using the same kind of figure of merit as in Eq. \ref{eq:connectorcleaning:kappa}. 
For cluster connection, the $\theta$ and $d$ parameter are computed using the cluster properties : the position of the cluster center of gravity (COG) and the direction of the cluster principal axis. 
$\theta$ is the angle between the two cluster principal axes and $d$ is the distance of closest approach between the two axes. 
When one of the clusters contains less than a few hits, it is called a fragment. When considering a fragment, $d$ is taken as the distance between the two COG and $\theta$ as the angle between the two COG directions from the detector center.

Neutral and charged clusters linked at the end of the selection are merged if the difference between the total merged cluster energy and 
the track momentum stay compatible with the calorimeter resolution. 

\section{Results}
\label{sec:results}

The PFA has been tested on the ILD option 2 in its original size design with simulations of $e^{+}e^{-} \to q\bar{q}$ events where
$q = u, d, s$ and $|\cos\theta_q| < 0.7$ (jets in the barrel). 
The jet energy resolution is computed as:
\begin{equation}
\label{eq:jetresol}
\frac{\text{RMS}_{90}(\text{E}_j)}{\text{Mean}_{90}(\text{E}_j)}
= \sqrt{2} \cdot \frac{\mathrm{RMS}_{90}(\mathrm{E}_{jj})}{\mathrm{Mean}_{90}(\mathrm{E}_{jj})},
\end{equation}
where $\mathrm{Mean}_{90}(\mathrm{E}_{jj})$ and $\mathrm{RMS}_{90}(\mathrm{E}_{jj})$ are the mean and RMS of the distribution of the total event energy computed in the smallest interval containing 90\% of the simulated events. 

For a center of mass energy of 91.2~GeV, the obtained resolution is 4.2\% for APRIL, 4.1\% for PandoraPFA and 3.25\% for Pandora's perfect PFA.
The perfect PFA is the reconstruction done by using Monte Carlo information to assign hits to the right particle.
The main contributions to the jet energy resolution in APRIL still come from the cluster merging and from the final track-cluster association.

\begin{figure}[htbp]
\centering 
\includegraphics[width=1.\textwidth]{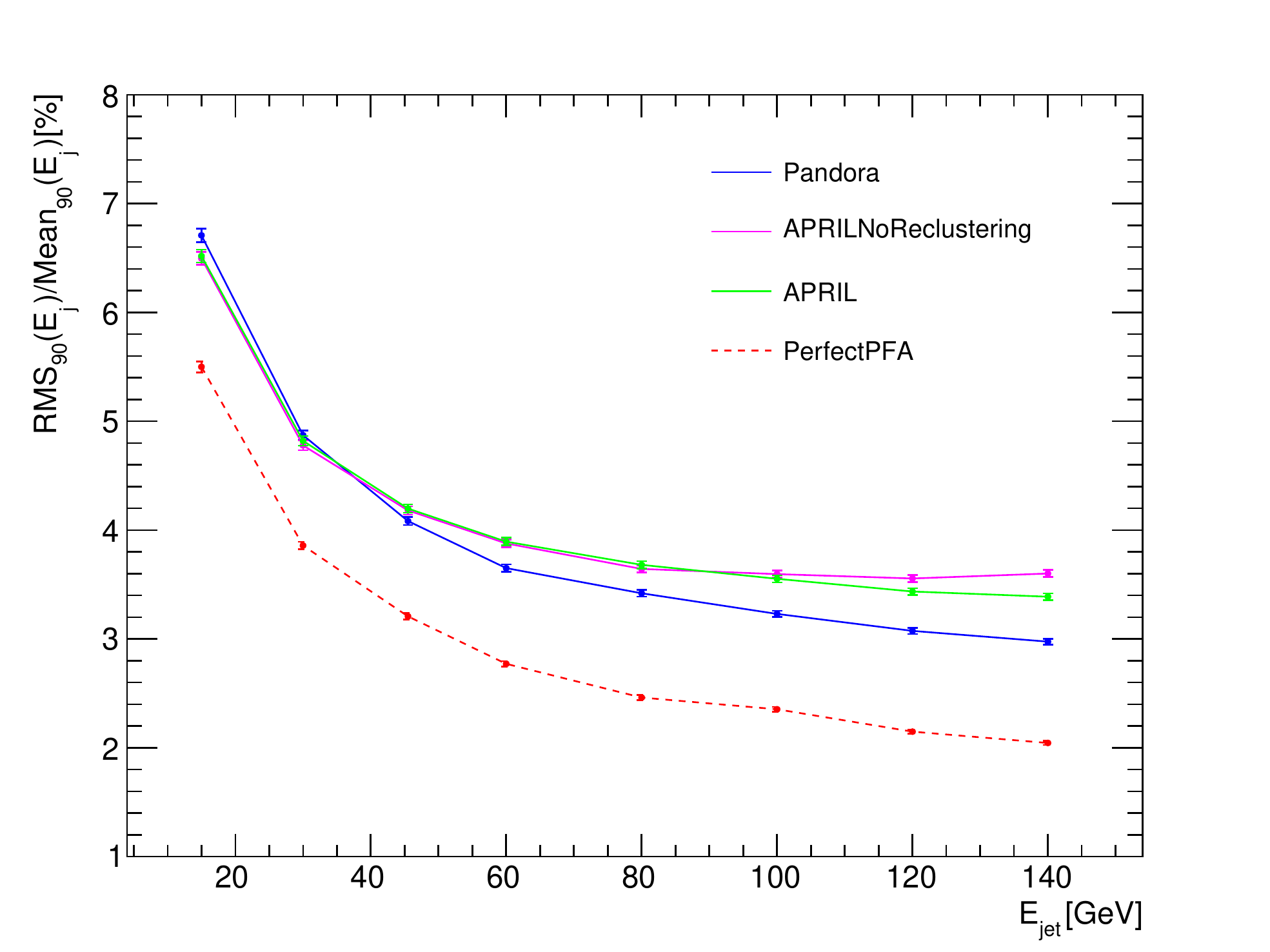}
\caption{\label{fig:iv} Jet energy resolution for various PFA reconstruction in dijets events as a function of the jet energy.}
\end{figure}

Figure \ref{fig:iv} shows the jet energy resolution obtained with APRIL and PandoraPFA as a function of the jet energy. 
At lower jet energy, APRIL is doing slightly better than Pandora. At higher jet energy, Pandora is doing better.
This is due to the reclustering method in Pandora. The reclustering method consists in redoing completely the PFA with different parameters on the portion of the events where the cluster energy and the corresponding track momentum mismatch. The reclustering chooses the PFA parameters set with the best match.
A simple reclustering has been tried in APRIL: when the cluster energy is too high compared to the track momentum, hits are removed from the charged cluster until the cluster energy matches the track momentum. Removed clusters are grouped in a neutral cluster. With this rough reclustering, the APRIL jet energy resolution is improved at high energy, indicating that implementation of a full reclustering scheme in APRIL might lead to an improved performance.

A PFA implementing the ARBOR approach has been developed in the PandoraSDK framework. At low and intermediate jet energies, it matches PandoraPFA performance for the SDHCAL option of the ILD. Implementation of a cluster splitting and/or reclustering procedure is expected to improve the reconstruction at higher jet energies. On a longer term, the APRIL algorithm will be adapted to deal with the non-linear energy reconstruction scheme used in the SDHCAL test beam data.


\acknowledgments

We are grateful to the LABEX Lyon Institute of Origins (ANR-10-LABX-0066) of the Universit\'e de Lyon for its financial support within the program "Investissements d'Avenir" (ANR-11-IDEX-0007) of the French government operated by the National Research Agency (ANR). 
This work was also in part supported by the AIDA-2020 project.


\end{document}